\documentclass{revtex4}  
\usepackage[latin1]{inputenc}
\usepackage{amsmath}
\usepackage{amsfonts}
\usepackage{amssymb}
\usepackage{graphicx,color}

\begin{document}

\title{Temporal ghost imaging with pseudo-thermal speckle light}

\author{Fabrice Devaux}
\email[Corresponding author: ]{fabrice.devaux@univ-fcomte.fr}
\affiliation{Institut FEMTO-ST, D\'epartement d'Optique P. M. Duffieux, UMR 6174 CNRS \\ Universit\'e Bourgogne Franche-Comt\'e, 15b Avenue des Montboucons, 25030 Besan\c{c}on - France}
\author{Kien Phan Huy}
\affiliation{Institut FEMTO-ST, D\'epartement d'Optique P. M. Duffieux, UMR 6174 CNRS \\ Universit\'e Bourgogne Franche-Comt\'e, 15b Avenue des Montboucons, 25030 Besan\c{c}on - France}
\author{Paul-Antoine Moreau}
\altaffiliation[Present address: ]{Centre for Quantum Photonics, H. H. Wills Physics Laboratory and Department of Electrical and Electronic Engineering, University of Bristol, Merchant Venturers Building, Woodland Road, Bristol BS8 1UB, United Kingdom}
\affiliation{Institut FEMTO-ST, D\'epartement d'Optique P. M. Duffieux, UMR 6174 CNRS \\ Universit\'e Bourgogne Franche-Comt\'e, 15b Avenue des Montboucons, 25030 Besan\c{c}on - France}

\author{S\'everine Denis}
\affiliation{Institut FEMTO-ST, D\'epartement d'Optique P. M. Duffieux, UMR 6174 CNRS \\ Universit\'e Bourgogne Franche-Comt\'e, 15b Avenue des Montboucons, 25030 Besan\c{c}on - France}
\author{Eric Lantz}
\affiliation{Institut FEMTO-ST, D\'epartement d'Optique P. M. Duffieux, UMR 6174 CNRS \\ Universit\'e Bourgogne Franche-Comt\'e, 15b Avenue des Montboucons, 25030 Besan\c{c}on - France}

\date{\today}

\begin{abstract}
We report ghost imaging of a single non-reproducible temporal signal in the range of tens $kHz$ by using pseudo-thermal speckle light patterns and a single detector array with a million of pixels working without any temporal resolution. A set of speckle patterns is generated deterministically at radio-frequency rate, multiplied by the temporal signal and time integrated in a single shot by the camera. The temporal information is retrieved by computing the spatial intensity correlations between this time integrated image and each speckle pattern of the set. 
\end{abstract}

\maketitle
\section{Introduction}
With the advent of laser sources in the early 1960s and their application in many research fields, some phenomena inherent to coherency of lasers were evidenced such as the speckle phenomenon \cite{allen_analysis_1963}. Very quickly, the scientific community investigated this phenomenon \cite{goodman_probability_1975} either to minimize its effects or in order to take advantage of its properties for different metrology applications \cite{dainty_1975}. Since then, speckle applications have increased and thanks to constant advances in technology, new developments are proposed \cite{goodman_speckle_2007,kaufmann_2011}.

Thus, exploitation of the statistical properties of speckle is the cause of fascinating new applications such as ghost imaging which is a way to form images of an object with a Single Point Detector (SPD) that does not have spatial resolution. The initial works used the quantum nature of entanglement of a two-photons state, where photons of a pair are spatially correlated, to detect temporal coincidences \cite{pittman_optical_1995}. While one of the photons passing through the object was detected by a photon counter with no spatial resolution, its twin photon was detected with spatial resolution by scanning the transverse plane with a single detector. Then, the object image was reconstructed by computing the correlation between both detectors. Later, similar experiments exploiting the temporal correlations of the intensity fluctuations of classical light \cite{bennink_two-photon_2002} or pseudo-thermal speckle light \cite{ferri_high-resolution_2005} were proposed, showing that entanglement was not necessary to ghost imaging.

By taking into account space-time duality in optics, the extension of the results of spatial ghost imaging to the time domain has been investigated with different kinds of light source \cite{shirai_temporal_2010,setala_fractional_2010,cho_temporal_2012,chen_temporal_2013,ryczkowski_ghost_2016}. For all proposed arrangements, the light emitted by the sources was split into two arms, called "reference" and "test" arms. While in the test arm the light was transmitted through a "time object" and detected with a slow SPD that can not properly resolve the time object, in the reference arm the light, that did not interact with the temporal object, was detected with a fast SPD. As for spatial ghost imaging, the temporal object was reconstructed by measuring the correlations of the temporal intensity fluctuations or the temporal coincidences between the two arms. In \cite{ryczkowski_ghost_2016}, measurements over several thousands copies of the temporal signal were necessary to retrieve an embedded binary signal with a good signal-to-noise ratio. The extension of spatial ghost imaging to the time domain looks attractive for dynamic imaging of ultra-fast waveforms with high resolution. However, the method used in \cite{ryczkowski_ghost_2016} requires many realizations of the same temporal signal, limiting the current applications to the detection of synchronized and reproducible signals \cite{faccio_optical_2016}. This is in contrast with spatial ghost imaging, where the object is unique, but multiplied in the time domain by a random modulation, different from one pixel to another, leading to multiplexing in this time domain.

Recently, we have proposed a different scheme which is the exact space-time transposition of computational ghost imaging \cite{shapiro_computational_2008} and of wavelength-multiplexing ghost imaging \cite{zhang_wavelength-multiplexing_2015}. In this experiment, a single complex temporal signal is measured with a single shot acquisition of a detector array without any temporal resolution \cite{devaux_computational_2016}. To achieve this, the temporal information is multiplexed in the spatial domain by multiplying it with computer-generated random images. As a result the temporal information is copied and transferred to the spatial domain. As the single shot is taken with a long exposure time, all the images are integrated by the camera. Although the temporal information seems to be lost, it is embedded in the time integrated image and it is retrieved by computing spatial correlations between the time integrated image and the computer-generated random images. This operation transfers the information back from the spatial domain to the temporal one.
We emphasise that this method allows to perform temporal ghost imaging without synchronization or replication of the temporal signal \cite{ryczkowski_ghost_2016} with a single detector array with thousands pixels that has no temporal resolution. Indeed, as for original computational ghost imaging arrangement \cite{shapiro_computational_2008}, the use of pre-computing and reproducible random binary patterns provides us to use a second array detector with a temporal resolution. Experiments were performed with monochrome images or color images to reconstruct either a single temporal signal or wavelength-multiplexed temporal signals, respectively. Currently, the main drawback of our method is its slowness (sampling rate of few $Hz$) which was imposed by the devices used to display the random patterns and to generate the temporal signals. 

In the present paper, we report the first experimental demonstration of temporal ghost imaging with pseudo-thermal speckle patterns that allows the retrieval of a single non-reproducible temporal signal at the $kHz$ rate. The signal reconstruction is performed by a single shot, spatially multiplexed, measurement of the temporal fluctuation of spatial intensity correlations of speckle patterns generated at tens of $kHz$ and modulated by the temporal signal to detect. Like in \cite{devaux_computational_2016}, and because speckle patterns are generated deterministically, images are recorded and time integrated with a single detector array with a million of pixels used without any temporal resolution.

\section{Experimental setup and measurement protocol}
\begin{figure}[htbp]
\centering
\fbox{\includegraphics[width=8cm]{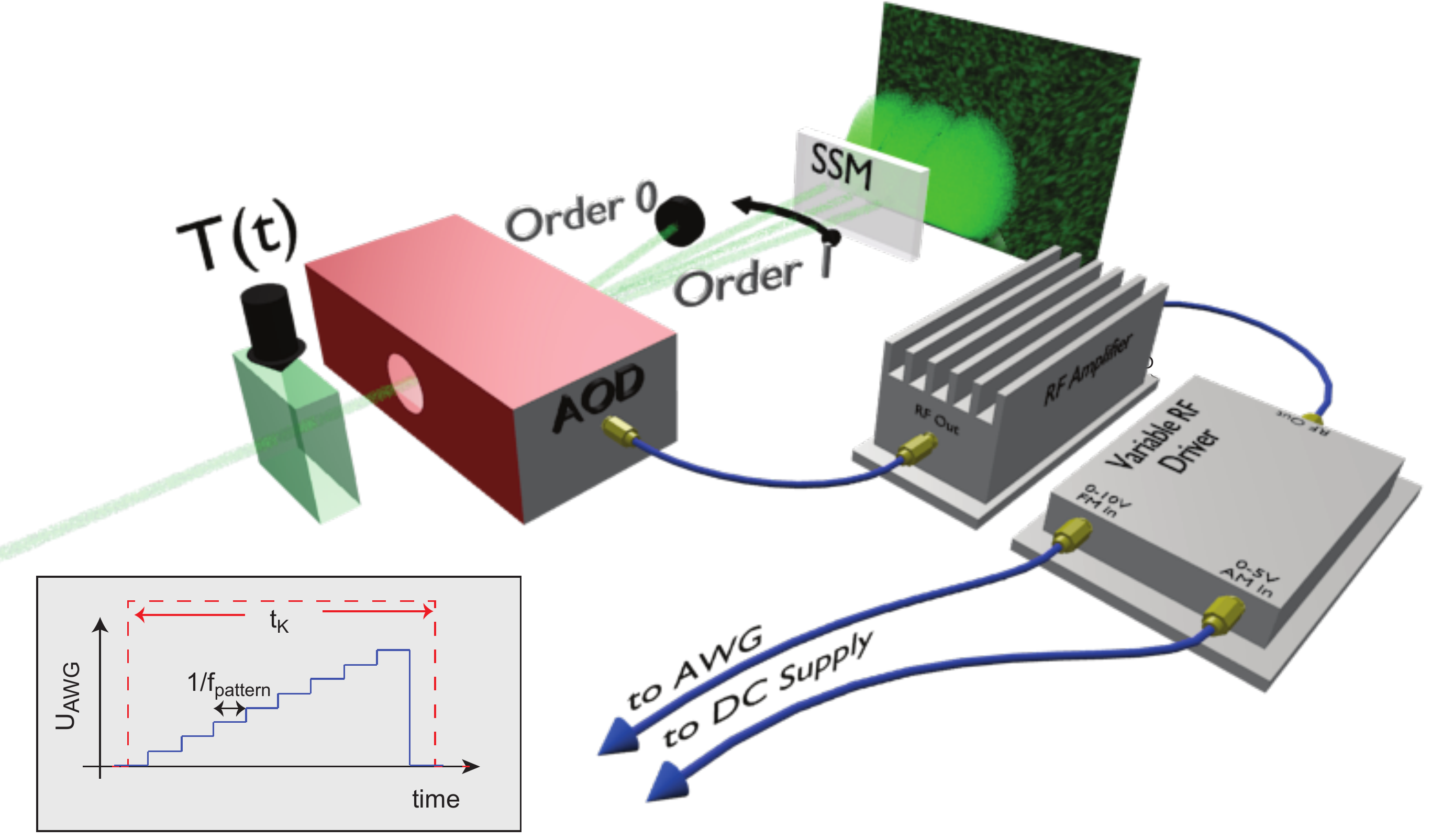}}
\caption{Experimental setup : A static scattering medium (SSM) is illuminated with a 543 $nm$ CW He-Ne laser beam. The basis of independent and reproducible speckle patterns is obtained by scanning the SSM with the laser beam by means of an acousto-optic deflector (AOD) controlled by a variable radio-frequency (RF) driver. The RF carrier (i.e. the deflection angle of the first diffraction order) is adjusted through the 0-10 $V$ input with a signal delivered by a programmable arbitrary wave generator (AWG) which is controlled by a graphical user interface (GUI). The insert shows a typical signal delivered by the AWG to generate successively 8 speckle patterns and the red dashed curve represents the duration $t_K$ of the signal delivered by the AWG. The temporal signal $T(t)$ is symbolized by a device that modulates the intensity of the laser beam.}
\label{setup}
\end{figure}

Fig. \ref{setup} illustrates the experimental setup. To generate deterministically a basis of speckle patterns, a 543 $nm$ CW He-Ne laser beam illuminates and scans the surface of a thin static scattering medium (SSM) by means of an acousto-optic deflector (AOD). The AOD is used in the Bragg configuration that gives a single, first diffraction order output beam, whose intensity and deflection angle are directly controlled with a variable radio-frequency (RF) driver. With our device, the RF carrier can be tuned in the range 40-100 $MHz$ by an analog signal through the frequency modulation input (0-10 $V$ FM in) which allows a deflection angle amplitude of 48 $mrad$ in random access or raster scan mode. In order to control the deflection angle accurately at a $kHz$ rate, the analog signal is delivered by a programmable arbitrary wave generator (AWG) linked to a computer with a GPIB cable and controlled by a graphical user interface (GUI). The intensity of the first diffraction order is controlled with a DC supply connected to the 0-5 $V$ amplitude modulation input (0-5 $V$ AM in). Images of the speckle patterns are acquired with a compact CMOS USB2.0 camera (IDS UI-1640C, 1280$\times$1024 pixels) on a 8 bits grey scale. We emphasise that our arrangement requires a single detector array because the scattering medium is static. So, the speckle patterns can be generated deterministically and reproduced with accuracy.    

Because of the laser beam divergence, the tuning of the deflection angle with a voltage step greater than or equal to $\Delta U_{AWG}=0.2\,V$ allows the generation of fully uncorrelated speckle patterns, providing the basis of independent random patterns required for the temporal ghost imaging protocol. This  protocol is similar to that described in \cite{devaux_computational_2016}, except that the basis of $K$ computed-random binary patterns is replaced by a basis of $K$ speckle patterns. Let us name  $X$ this basis.

\begin{figure}[htbp]
\centering
\fbox{\includegraphics[width=8cm]{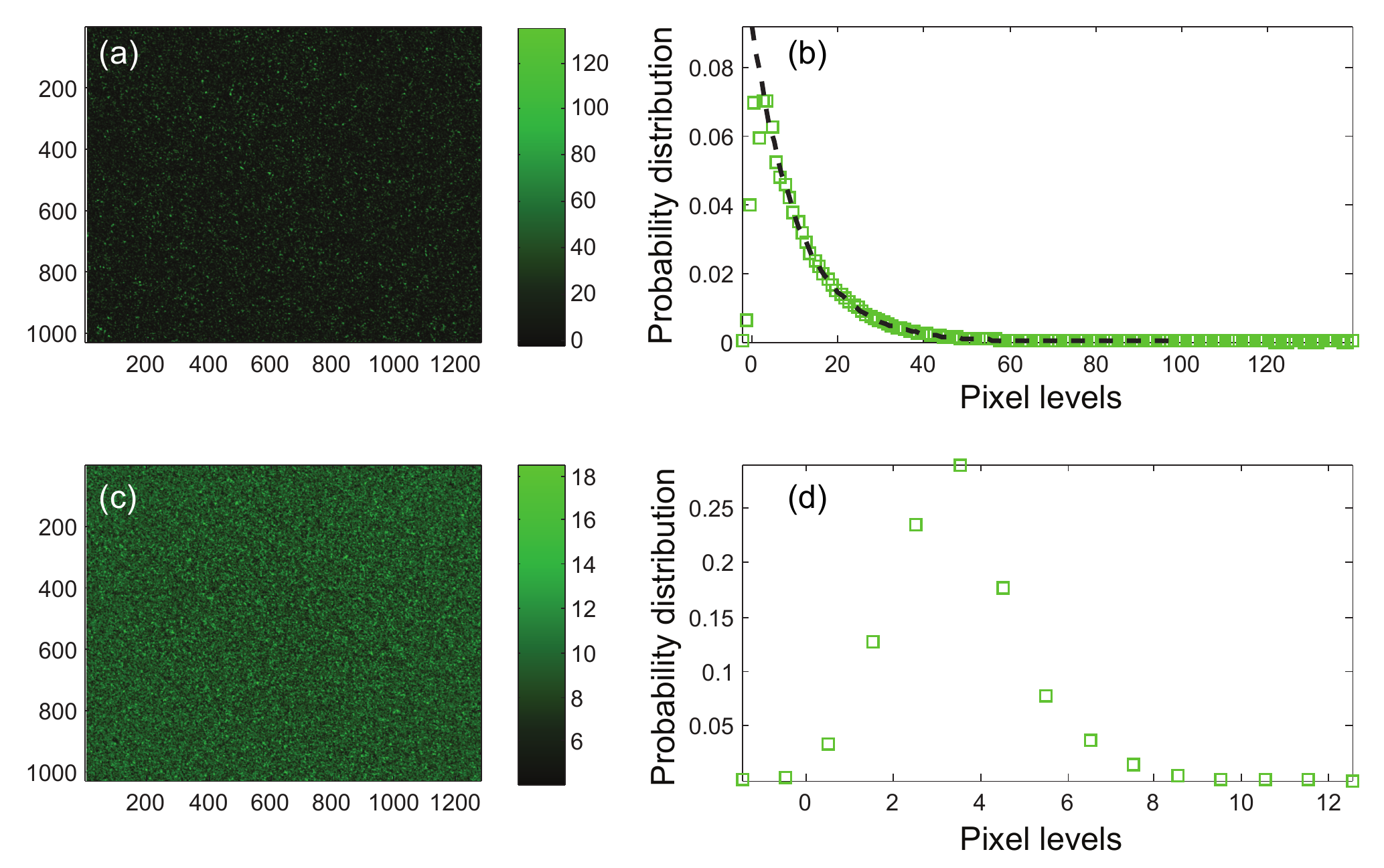}}
\caption{Images and distribution density of the pixel levels of: (a-b) a single speckle pattern, (c-d) a ghost image where a temporal signal is embedded. The dashed black curve in (b) corresponds to the exponential probability function of mean 11.}
\label{results:speckles}
\end{figure}

The first step of the protocol consists in taking with the camera an image of each speckle pattern generated for a specific deflection angle. Fig. \ref{results:speckles}a shows a typical image of a single speckle pattern and Fig. \ref{results:speckles}b shows the corresponding probability distribution of the pixel levels when the mean level of the background noise is subtracted. From these images, the mean and the variance of the speckles are calculated: $\langle X_k\rangle =11\pm 2$ and $\sigma^2_{X_k}=122\pm 32$. These values give a contrast ratio of the speckle patterns of $1.01\pm 0.04$ and the good agreement of the probability distribution of the pixel levels with an exponential function  (black dashed curve in Fig. \ref{results:speckles}b) shows that the statistical properties of speckles correspond to fully resolved pseudo-thermal light \cite{goodman_speckle_2007}. Moreover we have verified that the speckle patterns of the basis are independent and reproducible with a very good accuracy.

\begin{figure}[htbp]
\centering
\fbox{\includegraphics[width=8cm]{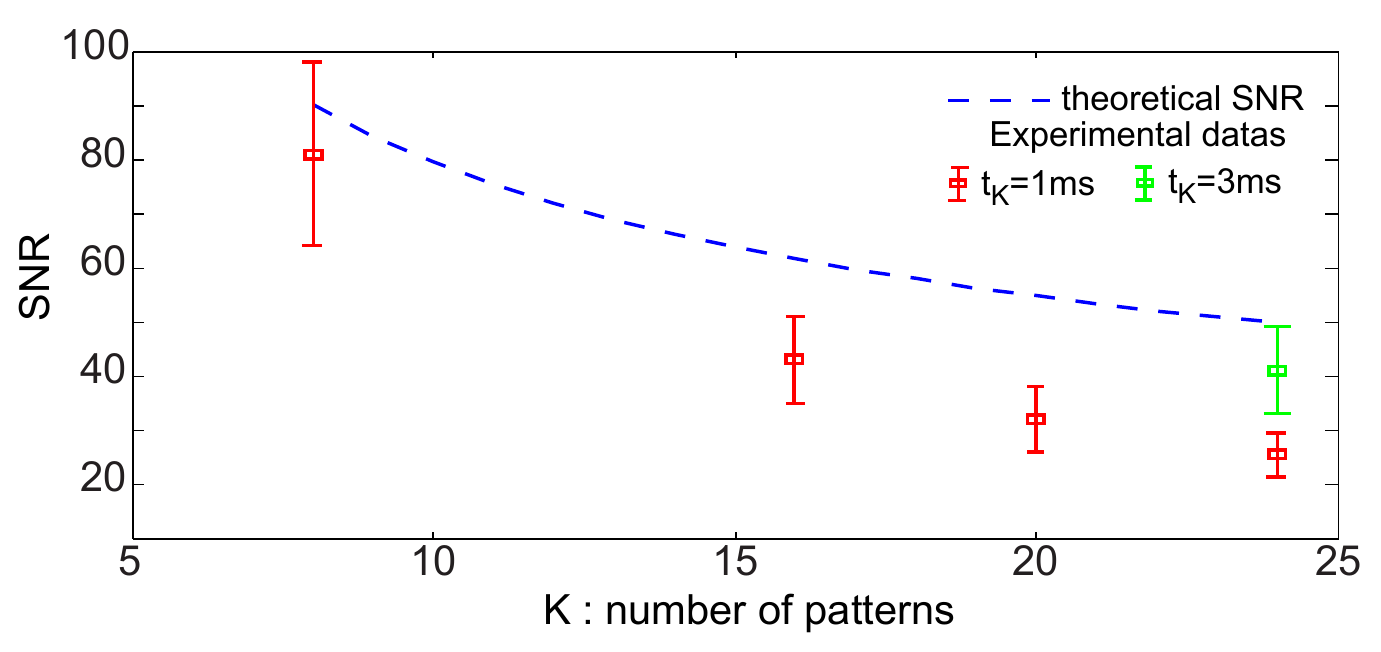}}
\caption{$SNR$ as a function of the number $K$ of time-intergrated speckle patterns. The blue dashed curve corresponds to the theoretical $SNR_{th}$ calculated with Eq. \ref{eq3} and experimental data are represented by the red and green squares. They respectively correspond to $1\,ms$ and $3\,ms$ durations of the voltage of the pulses trains delivered by the AWG. The error bars are deduced form measurements.}
\label{results:SNR}
\end{figure}

In a second step, the $K$ speckle patterns of the basis are sequentially generated by applying to the AOD a unique step by step increasing voltage delivered by the AWG at the frequency $f_{pattern}$ during a time $t_{K}$ (see insert in Fig. \ref{setup}). During this time, the temporal signal, named $T(t)$ ($0\leq T(t)\leq 1$) modulates the intensity of the laser beam and the intensity modulated patterns are time-integrated by the camera with an exposure time $t_{cam}\gg t_{K}$. 
Fig. \ref{results:speckles}c shows a typical time-integrated image with an embedded temporal signal ($t_{cam}=500\,ms$ , $t_K=1\,ms$, $K=24$ and $f_{pattern}=\frac{K}{t_K}=24\,kHz$). Fig. \ref{results:speckles}d shows the corresponding distribution of the pixel levels. The  mean level of the time integrated images is smaller than the mean level of the single speckle images because the SSM is illuminated with a longer time for acquisition of the individual speckle patterns.      
The time-integrated image, named $S$, verifies: 
\begin{equation}\label{eq1}
S_{ij}=\frac{1}{\gamma} \sum\limits_{k=1}^{K}T(k)X_{k,ij}
\end{equation}  
where $S_{ij}$ and $X_{k,ij}$ are the levels of the pixel of coordinates $(i,j)$ in the images $S$ and $X_k$, respectively. $T(k)$ is the value of the temporal signal at the time when the $k^{th}$ speckle is generated. $\gamma$ is a normalisation coefficient corresponding to the ratio between the illumination durations of the SSM when the $X_k$ and $S$ images are recorded. As in Ref. \cite{devaux_computational_2016}, the temporal signal is reconstructed by calculating the intensity correlations between the time integrated image $S$ and the $K$ images $X$ of the speckle patterns. The value of $T(k_0)$ at the "time" $k_0$ is estimated by (a hat means "estimator of"):
\begin{equation}\label{eq2}
\hat{T}(k_0)=\gamma \frac{\sum\limits^{N_x}_{i=1}\sum\limits^{N_y}_{j=1} \left(S_{ij}-\overline{S}\right)\left(X_{k_0,ij}-\overline{X}_{k_0}\right)}{\sum\limits^{N_x}_{i=1}\sum\limits^{N_y}_{j=1}\left(X_{k_0,ij}-\overline{X}_{k_0}\right)^2}
\end{equation}
where $\overline{S}$ and $\overline{X}_{k_0}$ are the arithmetic mean levels of the related images. $N_x$ and $N_y$ are the numbers of pixels in the images along the $x$ and $y$ dimensions. With numerical simulations, we verified that the signal-to-noise ratio ($SNR$) is given theoretically by (see supplement material of \cite{devaux_computational_2016}):
\begin{equation}\label{eq3}
SNR[T(k_0)]=\frac{T(k_0)}{\sigma_{T(k_0)}}=\sqrt{\frac{N_{eff}}{\sum\limits^K_{k=1,k\neq k_0}T^2(k)}}T(k_0) \geq \sqrt{\frac{N_{eff}}{K-1}}T(k_0)
\end{equation}  
where $N_{eff}$ represents the total number of independent spatial modes in the recorded image of a single speckle pattern. Typically, in Fig. \ref{results:speckles}a where a spatial mode covers an area of $23\pm 1$ pixels of the camera : $N_{eff}\backsimeq 5.7\times10^4$. For $K=24$ and $T=1$, it gives a theoretical $SNR_{th}= 50\pm 1$.
To measure the $SNR$, we applied the protocol described above and speckle patterns were time-integrated by the camera with $T=1$ during the entire exposure time. The protocol was repeated for different values of $K$ and $t_{K}=1\,ms$. Then, the transmission coefficient $T$ was calculated using Eq. \ref{eq2} and the $SNR$ was deduced from the fluctuations of the measured values of $T$. This measure was repeated several times in order to estimate the uncertainties. Fig. \ref{results:SNR} presents the measured $SNR$ as a function of $K$ when $t_{K}=1\,ms$ (red square) and error bars are deduced from measurements. To compare, the blue dashed curve represents the theoretical $SNR_{th}$ calculated with Eq. \ref{eq3}. We can observe that the measured $SNR$ becomes much smaller than the theoretical value when $K$ increases. Indeed, when $K$ increases, $f_{pattern}$ increases too and the $6.5\,\mu s$ access time of the AOD becomes of the same magnitude as the illumination time of the SSM when a speckle pattern is generated. For example, with $f_{pattern}=24\,kHz$, the laser beam illuminates the SSM during $42\,\mu s$ to generate a speckle pattern. However, the transition time between the successive positions of the laser beam represents a quarter of this duration. Consequently, this motion of the laser beam adds background noise that deteriorates the statistical properties of the speckle patterns and the $SNR$. The same protocol has been performed with $K=24$ and $t_K=3\,ms$ ($f_{pattern}=8\,kHz$). In that case, we measure a $SNR$ of $41\pm 8$ (green square in Fig. \ref{results:SNR}) which is in better agreement with the theoretical value.
 
\begin{figure}[htbp]
\centering
\fbox{\includegraphics[width=8cm]{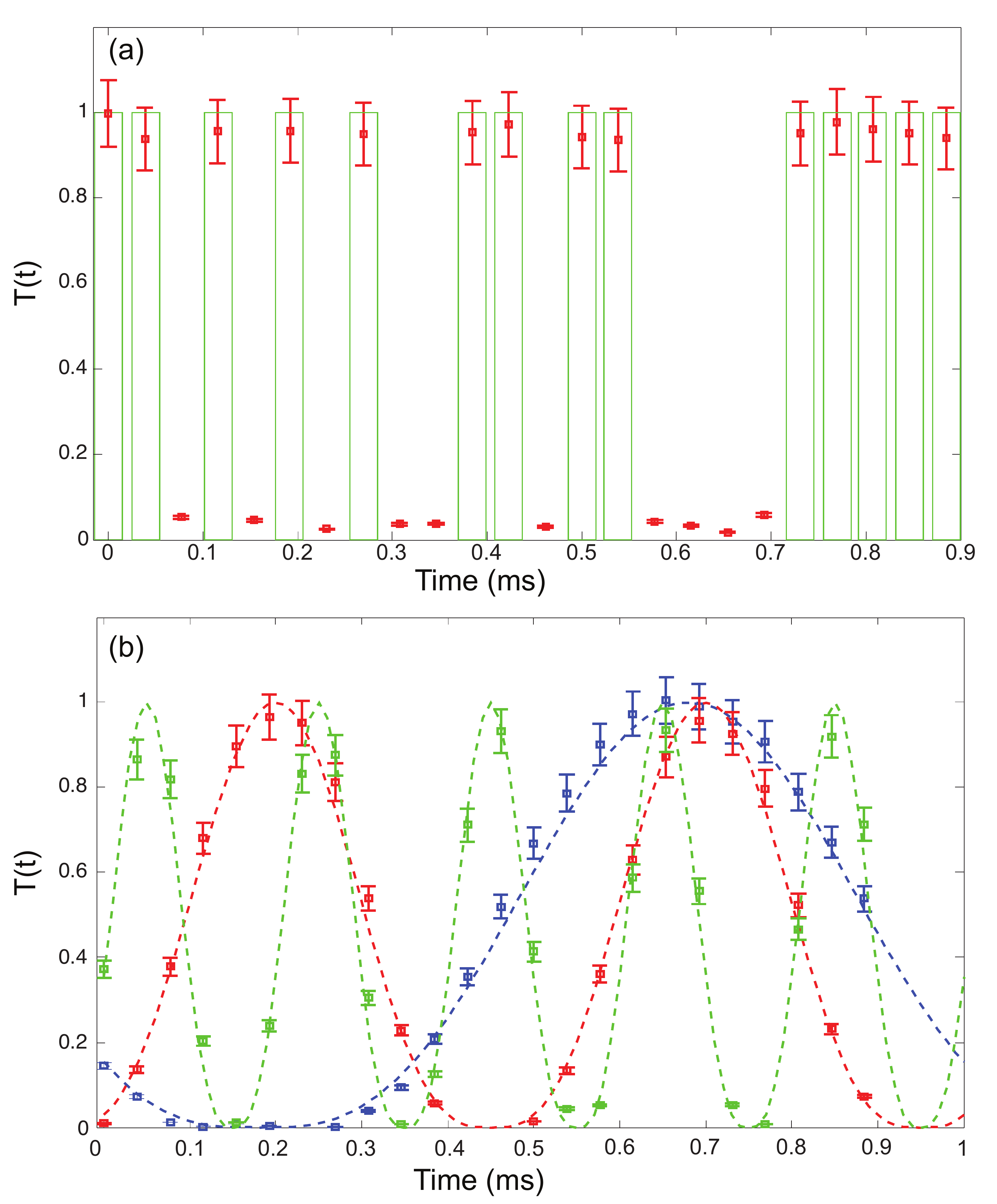}}
\caption{Results: (a) reconstructed random binary word of 24 bits at $f_{pattern}=24\,kHz$. Green bars show the original signals. (b) reconstructed periodic signals of frequencies 1, 2 and 5 $kHz$ (respectively blue, red and green squares) sampled at $f_{pattern}=24\,kHz$. The error bars are deduced from the measured $SNR$ and the dotted curves show the original signals. }
\label{results:reconstructedsignal}
\end{figure}

We present now the experimental results. Figures \ref{results:reconstructedsignal}(a) and \ref{results:reconstructedsignal}(b) show two kinds of reconstructed signals. The first one is a random binary word of 24 bits generated at the frequency $f_{pattern}=24\,kHz$ with the GUI interface. This binary signal modulates directly the amplitude of the pulses train delivered by the AWG. When the $k^{th}$ bit equals "1", the $k^{th}$ speckle pattern is generated with the corresponding voltage $U_{AWG}(k)$ and when it equals "0", $U_{AWG}(k)=0$ and the corresponding speckle pattern is not generated. Indeed, when $U_{AWG}=0\,V$, the first order diffracted beam is spatially filtered and does not illuminate the SSM. Then, temporal fluctuations of the spatial correlations between the time-integrated image and the 24 images of the speckle patterns basis are calculated with Eq. \ref{eq2}. In fig. \ref{results:reconstructedsignal}(a) we can observe that the original signal represented by the green bars is reconstructed experimentally (red squares) with a good accuracy. Note that the spatial correlations between the ghost image and the missing patterns (corresponding to bits "0") are almost null, which clearly shows the independence of the speckle patterns.
Fig. \ref{results:reconstructedsignal}(b) shows single shot measurements of non synchronised periodic signals of frequency 1,2 and 5 $kHz$ (respectively blue, red and green squares) sampled with the speckle patterns at the frequency $f_{pattern}=24\,kHz$. Here, the temporal signal is delivered by a wave function generator driving a device that modulates the transmission ($0\leq T(t)\leq 1$) of the laser beam  at the input of the AOD (see Fig. \ref{setup}). The original signals are represented by the dashed curves (only phases of these curves are adjusted to fit the experimental datas). For all experimental results the error bars are deduced from the measured $SNR$ when $f_{pattern}=24\,kHz$.

\section{Conclusion}
To summarize, these experiments represent the first demonstration of the exact space-time transposition of spatial ghost imaging with pseudo-thermal speckle light. Unique and non reproducible time objects are reconstructed with a very good accuracy by multiplying them with independent and reproducible speckle patterns with a very large number of independent spatial modes. This method ensures spatial multiplexing of temporal intensity correlations before detection of the time integrated images of the speckle patterns with a camera that has no temporal resolution. When compared to the recent experimental demonstrations of temporal ghost imaging, the main advantages of our system consist first in the replacement of the need of thousands synchronized replica of the temporal signal required in \cite{ryczkowski_ghost_2016} by the use of a detector array with a million pixels to acquire a single non reproducible temporal signal, second in a great improvement of the acquisition rate, from $Hz$ to tens of $kHz$, with respect to our previous device for computational temporal ghost imaging \cite{devaux_computational_2016}. Our results also demonstrate that deterministic speckle patterns with many spatial modes can be quickly controlled and addressed. Consequently, performances of our experiment, which are limited by the $6.5 \mu s$ access time of the AOD, could be seriously improved using spatial multiplexing of temporally modulated light sources, like 2D VCSELs array \cite{grabherr_vcsel_2014}, to illuminate the SSM with a rate up to 8 $GHz$. It would be also possible to retrieve other signals affecting the laser beam, like phase or wavelength, by modifying the first, learning step of the protocol.
  
\section*{Funding}
This work was supported by the Labex ACTION program (ANR-11-LABX-0001-01).

\bibliography{ThermalGhostImaging}

\end{document}